# Role of hydrogen dynamics and deposition conditions in photochromic YHO/MoO$_3$ bilayer films

E. Strods[1*], M. Zubkins[1], V. Vibornijs[1], D. Moldarev[2], A. Sarakovskis[1], K. Kundzins[1], E. Letko[1], D. Primetzhofer[2], J. Purans[1]

1. Institute of Solid State Physics, University of Latvia, Kengaraga 8, LV-1063, Riga, Latvia

2. Department of Physics and Astronomy, Ångström Laboratory, Uppsala University, Box 516, SE-751 20 Uppsala, Sweden

*Corresponding author: edvards.strods@cfi.lu.lv

## Abstract

Oxygen-containing yttrium hydride (YHO) and molybdenum trioxide (MoO$_3$) bilayer films (YHO/MoO$_3$) are produced using reactive magnetron sputtering, and their photochromic properties are investigated in relation to the thickness and density of the MoO$_3$ layer. Compared to single YHO films, the YHO/MoO$_3$ films exhibit faster coloration and larger contrast, with both parameters adjustable by varying the thickness or deposition pressure of the MoO$_3$ layer. Transparent YHO/MoO$_3$ films (~75% at 550 nm) demonstrate a photochromic contrast of up to 60%, significantly higher than the 25–30% contrast observed for single YHO films after 20 hours of UVA-violet light exposure. This enhancement arises from hydrogen intercalation from the (200)-textured polycrystalline YHO film into the X-ray amorphous MoO$_3$, leading to the formation of molybdenum bronze (H$_x$MoO$_3$), as confirmed by X-ray photoelectron and optical spectroscopies. However, the darkened YHO/MoO$_3$ films do not fully recover to their initial transparency after illumination due to the irreversible nature of the coloured MoO$_3$ layer. Most of the hydrogen intercalated into MoO$_3$ originates from the YHO layer during the initial darkening process. Furthermore, the bilayer films are chemically unstable, exhibiting gradual darkening over time even without intentional UV illumination, as confirmed by nuclear reaction analysis.

## 1. Introduction

The photochromic activity of oxygen-containing rare-earth (RE) hydride films (REHO) [1] depends on their composition [2], thickness [3], and illumination conditions [4]. It is suggested that an oxyhydride phase forms in these compounds, where both oxygen (O) and hydrogen (H) act as anions [5]. Mixed-anion compounds, such as oxyhydrides, constitute an intriguing class of materials with significant potential for future technologies [6], as they contain multiple anions, enabling the development of technologies with enhanced functionality. YHO layers with an appropriate O:H atomic ratio (~0.5–1.0 [7]) exhibit a neutral, reversible photochromic effect (with contrast up to approximately 50%) when illuminated with a solar simulator at room temperature and ambient pressure. The coloured films gradually revert to their initial transparent state in darkness, with bleaching times ranging from minutes to hours, strongly depending on the specific YHO films.

Despite increasing research on YHO films in recent years, the mechanisms underlying the observed photochromism and the precise phase of oxygen-containing rare-earth hydrides remain unresolved. Current studies propose several explanations for the photochromic behaviour in YHO films, including the formation of metallic domains due to either photo-excited electrons reducing trivalent yttrium [8] or anion diffusion [9–11]. The relatively long bleaching time suggests that the underlying process is not purely electronic. Indeed, in-situ studies of hydrogen loss from YHO films [12] indicate that mobile hydrogen plays a pivotal role in the photochromic effect, with photon-induced hydrogen diffusion between two separate phases in GdHO proposed in Ref [13]. Additional mechanisms, such as the formation of hydroxides, colour centres, or dihydrogen [14], are discussed in detail in Ref [15].

The photochromic YHO phase is obtained by exposing $\beta$-YH$_{2-x}$ films to an oxygen-containing atmosphere [16]. During this process, the films oxidise to a certain extent due to their high reactivity with oxygen, resulting in transparent and photochromic properties [17], with the film density, controlled by deposition pressure, being a critical parameter [18]. The oxidation process involves both hydrogen release and rearrangement, leading to changes in the electronic state [19]. In the $\beta$-YH$_{2-x}$ structure, all hydrogen atoms occupy tetrahedral positions. As the material oxidises, oxygen anions replace hydrogen in these tetrahedral positions, with some hydrogen relocating to octahedral positions [20,21]. This structural rearrangement has been confirmed through extended X-ray absorption fine structure (EXAFS) studies [22].

Recent studies have shown that YHO films coated with tungsten trioxide ($WO_3$) exhibit distinctly different photochromic behaviour [23,24]. These composites demonstrate faster coloration and greater solar modulation ($\Delta T_{sol}$ = 50%) compared to single YHO films, explained by hydrogen migration between the layers. Heating during the bleaching process further enables rapid optical switching. However, additional research is required to better understand the synergistic effects of hydrogen migration and the impact of chromic transition metal oxides, including their deposition parameters, on the photochromic properties of these composite films.

Molybdenum oxides are highly versatile materials, exhibiting a range of stoichiometries that support their use in diverse high-value research and commercial applications. Their oxidation states can be tailored to adjust crystal structure, morphology, oxygen vacancies, and dopants, allowing precise control over their optical and electronic properties [25]. Molybdenum trioxide ($MoO_3$), while sharing similarities with its structural analogue $WO_3$ [26], exhibits pronounced chromic behaviours, including electrochromism [27], gasochromism [28], photochromism [29], and thermochromism [30]. Notably, its electrochromic response exhibits stronger and more uniform light absorption in the coloured state compared to $WO_3$ [31]. This performance is particularly effective for human vision due to its intense spectral features at wavelengths corresponding to peak visual sensitivity. Various theoretical models, such as the colour centre model, intervalence-charge transfer (IVCT) model, and small-polaron absorption model, have been proposed to explain the photochromic behaviour of $MoO_3$ under UV light [32].

In this study, YHO/$MoO_3$ bilayer films are produced using reactive magnetron sputtering to examine the influence of $MoO_3$ thickness and deposition pressure on the photochromic response. Based on nuclear reaction analysis (NRA) and X-ray photoelectron spectroscopy (XPS), we propose that hydrogen is released from the YHO film, leading to the subsequent formation of molybdenum bronze ($H_xMoO_3$). The findings provide valuable insights into the photochromic mechanisms of YHO and suggest strategies for optimizing the optical and photochromic properties of the bilayer coatings, contributing to the advancement of future photochromic devices.

## 2. Experimental details

The thin film deposition was performed using a vacuum PVD coater, model G500M, manufactured by Sidrabe Vacuum, Ltd. $YH_{2-x}$ thin films were deposited onto soda-lime glass and undoped silicon (100) wafer (MicroChemicals GmbH) substrates by reactive pulsed direct current (pulsed-DC) magnetron sputtering (80 kHz, 2.5 µs off-time) using an Y (purity 99.9%) target in an Ar (purity 99.9999%) and $H_2$ (purity 99.999%) atmosphere, with a $H_2$:Ar gas flow ratio of 1:8. A planar balanced magnetron with target dimensions of 150 mm × 75 mm × 3 mm was utilized, positioned at a distance of 10 cm from the grounded substrate holder, with substrates aligned along the target axis. No intentional heating of the substrates was applied during the sputtering process. Prior to deposition, the chamber was evacuated to a base pressure of $6\times10^{-4}$ Pa using a turbo-molecular pump backed by a rotary pump. The yttrium target was sputtered at a constant average power of 200 W for 20 minutes, under a working pressure of 0.75 Pa. To achieve the YHO phase, air was deliberately introduced into the chamber immediately after deposition. Afterward, the chamber was again evacuated to the base pressure, and an YHO film was coated with $MoO_3$. An Mo target (purity 99.95%) was sputtered at a constant average power of 200 W in an Ar and $O_2$ (purity 99.999%) atmosphere, maintaining an $O_2$:Ar gas flow ratio of 1:3. Depending on the process, different deposition parameters were varied: deposition time in the range of 8 to 60 minutes to vary the thickness between 20 and 400 nm and deposition pressure in the range of 0.1 to 2.7 Pa. After the deposition of the $MoO_3$ film, air was introduced into the chamber, after which the samples were removed and cut into two pieces. One piece was stored in an inert (Ar) atmosphere, while the other was immediately analysed using UV-Vis-NIR spectroscopy, X-ray diffraction (XRD), and a photochromic response measurement system. The stored samples were later analysed by X-ray photoelectron spectroscopy (XPS), nuclear reaction analysis (NRA), and scanning electron microscopy (SEM). Additionally, single $MoO_3$ films were deposited on soda-lime glass and silicon substrates using the same parameter values (pressure and thickness) as those used in the previous sets of YHO/$MoO_3$ samples.

The crystallographic structure of the films was examined by XRD, using a Rigaku MiniFlex600 X-ray powder diffractometer with Bragg-Brentano θ-2θ geometry and a 600 W Cu anode (Cu Kα radiation, λ = 1.5406 Å) X-ray tube.

The surface and cross-section of the films was visualized using Helios 5 UX dual-beam scanning electron microscope (SEM) form Thermo Fisher Scientific. The cross-section images were obtained by scratching the film off the substrate and imaging scratched pieces on the substrate, which are placed closely perpendicular to the substrate. Surface SEM images were processed using local adaptive thresholding to segment and analyse the visible grain area.

Film thicknesses were determined using a SE WOOLLAM RC2 spectroscopic ellipsometer in the spectral range of 550–1690 nm (2.25–0.73 eV), where the films exhibit high transparency. The Cauchy model was applied in this range for the analysis. The primary ellipsometric angles, Ψ and Δ, were measured at incident angles ranging from 55° to 65° in 5° increments. Spectroscopic ellipsometry (SE) experimental data and model-based regression analyses were processed using the WOOLLAM software, CompleteEASE. The mean squared error (MSE) values for the models ranged between 1 and 20. The approximate thickness of the YHO films in this study ranged from 720 to 790 nm. A detailed analysis of the SE measurements over the broader spectral range of 5.9–0.7 eV, employing physical oscillator models, along with the optical constants, their gradients, and the photochromic properties of YHO films deposited under the described conditions, is provided in Ref. [17].

The photochromic response was tested using UVA-violet light from a 15 W (≈2.4 mW/cm$^2$) lamp under ambient conditions. The energy of the source light at the maximum intensity is 3.3 eV with the full-width at half maximum (FWHM) of 0.13 eV. A tungsten-halogen lamp with the power density of ≈3.3 mW/cm$^2$ was used for the probing. No detectable photo-darkening caused by the probing light is observed. The schematic of the measurement system and the spectra of the light sources used are provided in Fig. S1 (Supplementary Information - SI).

The optical transmittance and reflectance of the films in the range 500–2500 nm were measured by an Agilent Cary 7000 spectrophotometer. The sample was placed at an angle of 6° against the incident beam of P-polarized light, and the detector was placed at an angle of 180° behind the sample to measure the transmittance and at 12° in front of the sample to measure the specular reflectance.

XPS measurements were performed by an ESCALAB 250Xi (ThermoFisher) instrument to determine the chemical composition of $MoO_3$ and $YHO/MoO_3$ films and the chemical state of Mo species. An Al Kα X-ray tube with energy 1486 eV was used as an excitation source, the size of the analysed sample area was 650 × 100 μm, the angle between the analyser and the sample surface was 90°. An electron gun was used to perform charge compensation. The base pressure, with the charge neutralizer switched on during spectra acquisition, was better than 10$^{-7}$ mbar achieved by rotary and turbomolecular pumps. The films were gently sputter-cleaned by $Ar^+$ cluster gun with the ion energy 6.0 keV (150 Ar atoms) for 10 s before the measurements. The size of the cleaned area was 2 × 2 mm. The pass energy of the measured spectra was 20 eV. The calibration and linearity of the binding energy scale was confirmed by measuring the positions of Ag $3d_{5/2}$, Au $4f_{7/2}$ and Cu $2p_{3/2}$ to be at 368.21 eV, 83.93 eV and 932.58 eV, respectively. The FWHM of Au $4f_{7/2}$ peak was better than 0.58 eV. The binding energy spectra of all the samples were referenced against the adventitious carbon peak to appear at 284.8 eV.

Fourier Transform Infrared (FTIR) absorbance spectra were measured using a VERTEX 80v vacuum spectrometer in transmittance mode. The experiments were conducted over a spectral range of 30 to either 4000 or 7000 $cm^{-1}$, with the interferometer operating under both vacuum and air conditions at a resolution of 4 $cm^{-1}$. Uncoated Si was used as the background reference.

Hydrogen depth profiling was conducted using NRA at the Tandem Laboratory at Uppsala University [33]. The probing beam of energetic $^{15}N$ ions enables the $^{15}N(^1H, \alpha\gamma)^{12}C$ reaction with the H in the sample. This reaction features a resonance at an energy of the $^{15}N$ ions of 6.385 MeV in the lab frame of reference. The emitted γ-rays at the characteristic energy of 4.43 MeV were counted using a bismuth germinate detector. The primary energy of the beam was scanned in the range 6.4–7.6 MeV to profile the depth distribution of H. The chemical composition of YHO films was identified using time-of-flight energy elastic recoil detection analysis (ToF-E ERDA) employing 36 MeV $^{127}I$ primary beam. For more experimental details we refer to [34].

## 3. Results and discussion

### 3.1. Structure of YHO/MoO$_3$

The XRD patterns of YHO and YHO/MoO$_3$ films on glass, where the MoO$_3$ layer thickness varies between 20 and 285 nm and deposition pressure ranges from 0.15 to 2.70 Pa, are shown in Fig. 1 over a 2θ range of 5°–90°. The XRD patterns reveal multiple peaks at 29°, 33°, 48°, 57°, and 73°, corresponding to the cubic YHO lattice planes (111), (200), (220), (311), and (400), respectively. The intense (200) peak indicates textured YHO films, with a crystalline phase growing preferentially along the perpendicular axis to the substrate. Meanwhile, the MoO$_3$ films are X-ray amorphous. The XRD patterns of the single MoO$_3$ films are presented in Fig. S2 (SI). However, Raman spectroscopy reveals that the X-ray amorphous structures of MoO$_3$ [35] and WO$_3$ [36] films consist of small nanocrystals embedded within an amorphous matrix. The atomic-scale interface may allow limited cation exchange or interfacial alloying. For instance, sub-nanometre-scale Y-O-Mo linkages could potentially form. In Ref. [24], the formation of a Y$_2$W$_3$O$_{12}$ phase was indicated by XRD and high-resolution transmission electron microscopy at the YHO/WO$_3$ interface. However, in the case of YHO/MoO$_3$ films, no corresponding diffraction peaks of Y$_2$Mo$_3$O$_{12}$ are observed, suggesting the absence of a similar crystalline interfacial phase. In general, the room-temperature deposition conditions limit interfacial diffusion and phase mixing.

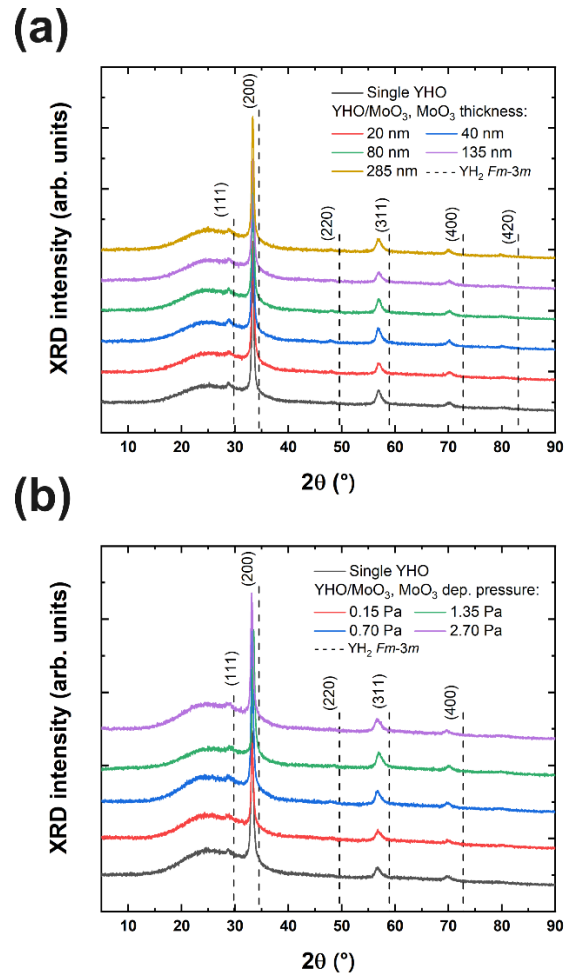

**Fig. 1.** XRD patterns of magnetron-sputtered YHO and YHO/MoO$_3$ films with MoO$_3$ (a) thickness varying between 20 and 285 nm and (b) deposition pressure ranging from 0.15 to 2.70 Pa. Vertical dashed lines indicate the Bragg peak positions for cubic $\beta$-YH$_2$ (ICDD 04-002-6939).

The XRD peaks are significantly shifted to lower angles compared to the $\beta$-YH$_2$ (*Fm-3m*) powder diffraction pattern (ICDD 04-002-6939), indicating lattice expansion due to oxygen incorporation, consistent with previous reports [37–39]. For all YHO and YHO/MoO$_3$ films, the lattice parameter ranges from 5.34 to 5.41 Å, while crystallite sizes are between 12.1 and 20.5 nm in the [200] direction calculated by the Scherrer equation.

The SEM surface images (Fig. 2(a,b)) show that YHO films with a thickness of 790 nm exhibit a fine-grained surface morphology. The features are densely packed, with relatively sharp edges and a mean area of approximately 5 000 nm$^2$ (Fig. S3). These surface features are composed of smaller crystallites (Fig. 2(b)), with sizes that closely match those determined by XRD. Voids between the surface features are clearly visible in YHO films covered with thinner MoO₃ layers of 40 and 80 nm, indicating incomplete coverage (Fig. 2(c,d)). These voids gradually decrease with increasing MoO₃ thickness, eventually resulting in a densely packed, fine-grained morphology once again (Fig. 2(f)). The mean area of surface features increases from approximately 7 000 to 14 000 nm² as the MoO₃ thickness increases from 40 to 285 nm (see Figs. S3 and S4). Cross-sectional images (Fig. 3) reveal that the YHO film grows in a columnar fibrous grain structure, with the columns broadening as growth progresses and vertical voids oriented perpendicular to the substrate surface, which are likely to act as fast hydrogen transport channels. Fig. 3(b-e) shows cross-sections of YHO/MoO₃ films with varying MoO₃ thicknesses, illustrating the structural evolution with increasing film thickness.

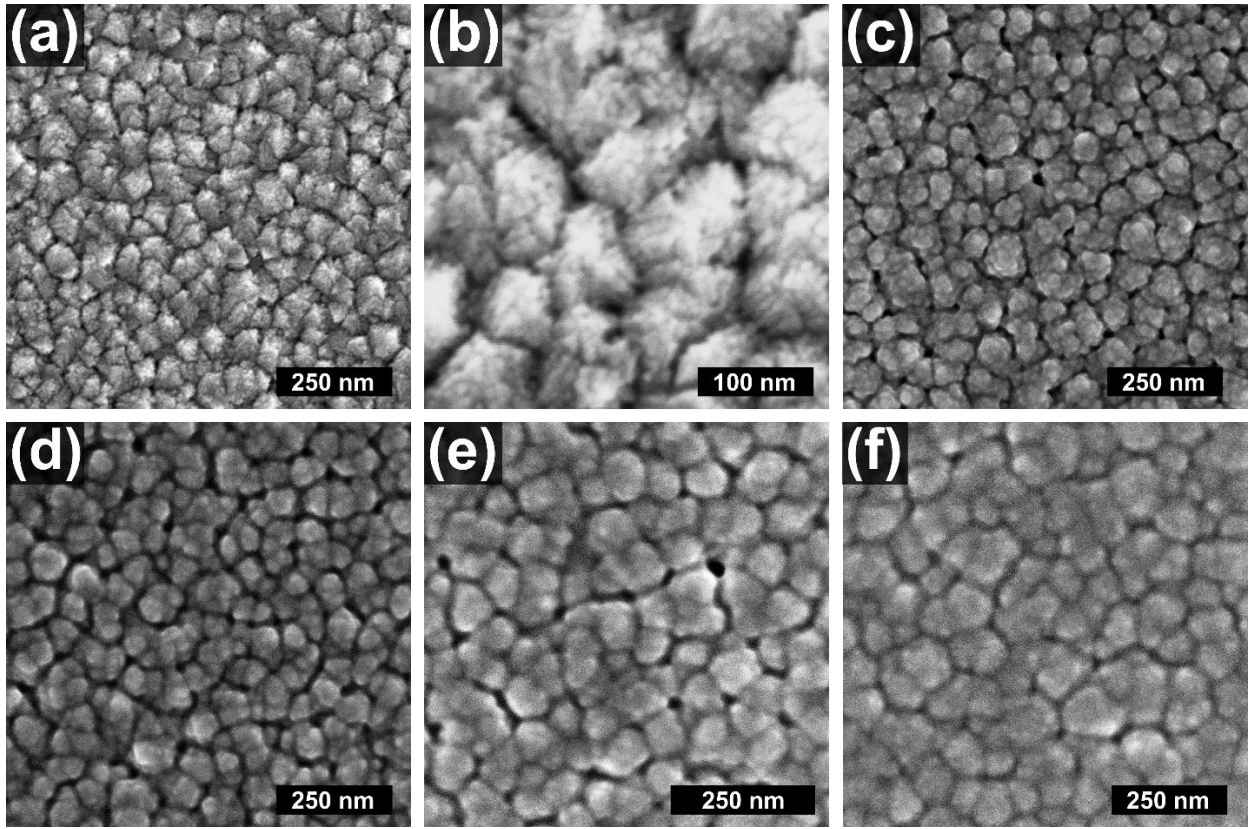

**Fig. 2.** Surface SEM images of magnetron-sputtered films. Images (a) and (b) show single YHO film captured at two different magnifications: (a) provides an overview of the surface morphology, while (b) presents a higher-magnification view that highlights the finer grain structure. Images (c)–(f) show YHO/MoO₃ bilayer films with MoO₃ thicknesses of (c) 40 nm, (d) 80 nm, (e) 135 nm,

and (f) 285 nm. All MoO₃ layers were deposited at a pressure of 1.35 Pa using an $O_2$-to-Ar gas flow ratio of 1:3.

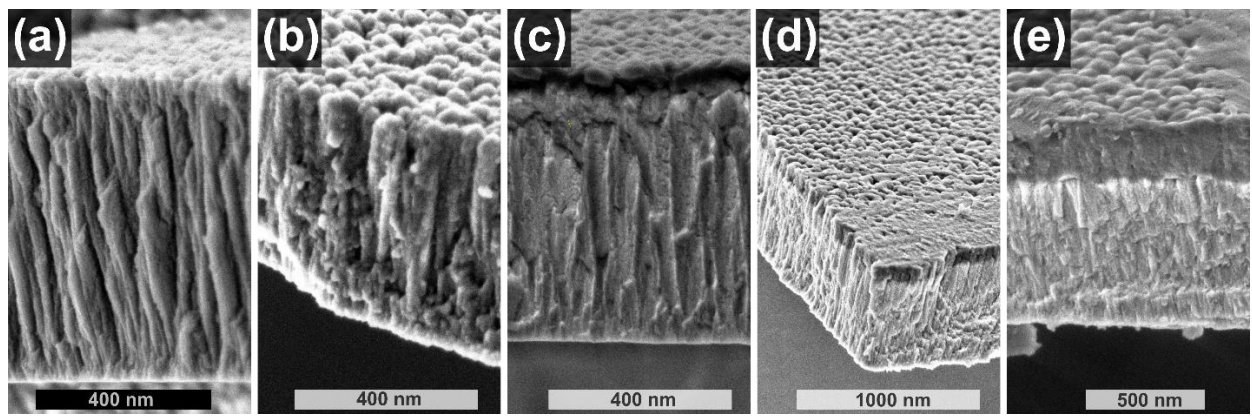

**Fig. 3.** Cross-sectional SEM images of magnetron-sputtered (a) single YHO and bilayer YHO/MoO₃ films with MoO₃ thicknesses of (b) 20 nm, (c) 80 nm, (d) 135 nm, and (e) 285 nm. The MoO₃ is deposited at a pressure of 1.35 Pa with an $O_2$-to-Ar gas flow ratio of 1:3.

*3.2. Optical and photochromic properties of YHO/MoO₃*

The light transmittance in the wavelength range of 250–2500 nm for non-illuminated and prolonged-illuminated (20 hours under UVA-violet light) YHO and YHO/MoO₃ films, with varying MoO₃ thicknesses and deposition pressures, is shown in Fig. 4. The oscillations in the spectra are caused by interference effects due to multiple reflections at the film interfaces. The sharp drop in transmittance below 400 nm corresponds to fundamental light absorption, occurring when the photon energy matches or exceeds the optical band gap. Both YHO and MoO₃ have optical band gaps within similar energy ranges – 2.7–3.6 eV [17,40] and 2.8–3.6 eV [41,42], respectively – depending on their chemical composition. The YHO and MoO₃ films deposited in this study exhibit optical band gaps of approximately 3.0 eV and 2.9–3.1 eV, respectively (see Tauc plots in Fig. S5 (SI)).

The initial transmittance of non-illuminated YHO and YHO/MoO₃ samples in the wavelength range of 500-2500 nm is approximately 75%. The transmittance spectra for non-illuminated YHO/MoO₃ samples are provided in Fig. S6 (SI). The photo-darkening effect is most pronounced in the visible light range and diminishes gradually in the infrared region. The spectra of prolonged-illuminated films clearly show an increased photochromic contrast for YHO/MoO$_3$ ($\Delta T$ = 55–60%) compared to YHO ($\Delta T$ = 25–30%) at 550 nm. Furthermore, for the YHO/MoO$_3$ film with a 285 nm MoO$_3$ layer, a broad absorption band in the range of 700–850 nm, characteristic of coloured MoO$_3$ [43], is observed. The photochromic contrast in YHO/MoO$_3$ films can be enhanced by increasing the MoO$_3$ film thickness or reducing the deposition pressure during MoO$_3$ growth. These changes and their underlying mechanisms are discussed in detail below.

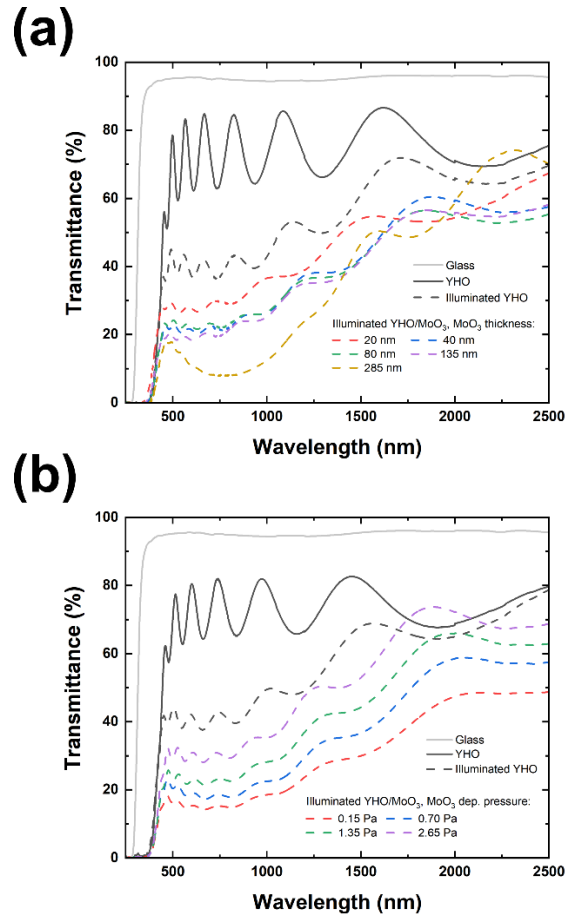

**Fig. 4.** Transmittance in the wavelength range of 250–2500 nm for clear and illuminated YHO films, and illuminated YHO/MoO$_3$ films (20 hours under UVA-violet light), with varying MoO$_3$ (a) thickness (deposition pressure of 1.35 Pa) and (b) deposition pressure (film thickness ≈80 nm).

Photo-darkening and bleaching measurements of the thin films were performed by illuminating the samples from the MoO$_3$ side using 3.3 eV UVA-violet radiation while monitoring changes in average light transmittance within the 500–700 nm range (Fig. 5). Each measurement cycle lasted two hours: one hour with the lamp on, followed by one hour with the lamp off.

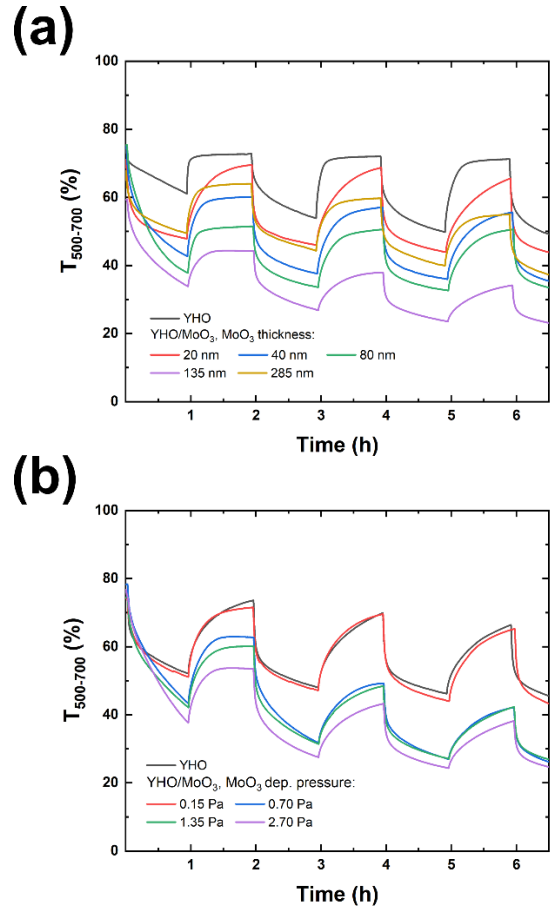

**Fig. 5.** Average transmittance between 500 and 700 nm during first three photo-darkening and bleaching cycles for magnetron-sputtered YHO and YHO/MoO$_3$ films with varying MoO$_3$ (a) thickness (deposition pressure of 1.35 Pa) and (b) deposition pressure (film thickness ≈80 nm).

The results indicate that MoO$_3$ films significantly enhance the photodarkening effect in the YHO/MoO$_3$ structure compared to single YHO films during the initial cycle. This enhancement is primarily attributed to the interaction between the two films, with MoO$_3$ acting as a complementary medium into which hydrogen can be injected from YHO. To fully understand the mechanisms underlying this phenomenon, it is essential to examine the distinct roles of each film and the processes occurring within them.

Firstly, the single MoO$_3$ films produced in this study do not exhibit inherent photochromic behaviour under UV illumination. XPS measurements indicate that the films have an O-to-Mo atomic % ratio ranging from 2.95 to 3.20, depending on the deposition pressure (see Table S1 in the SI), and exhibit a relatively weak water signal in the FTIR spectra (Fig. S7 in the SI). This observation aligns with previous reports on thin, amorphous MoO$_3$ films with a near-stoichiometric composition ($x < 0.3$ in MoO$_{3-x}$), where photochromism is typically low due to the lack of oxygen vacancies or interstitial water [44]. Although oxygen vacancies are known to play a crucial role in the chromogenic properties of MoO$_3$, their exact function remains incompletely understood [45]. Additionally, the absence of water eliminates the possibility of proton-mediated chromic reactions, such as those observed in other transition metal oxides like WO$_3$.

Secondly, a previous study [46] reported that molecular hydrogen is released during the initial darkening cycle of freshly deposited YHO films using 455 nm blue light, with negligible amounts detected in subsequent cycles. This observation was attributed to weakly bound hydrogen, most likely located at grain boundaries. Hydrogen has the ability, due to its small kinetic diameter, to deeply intercalate into the MoO$_3$ film. The injected hydrogen is expected to be oxidised under UV illumination, with electron-hole generation producing protons, given that MoO$_3$ is highly oxidative due to its strongly positive valence band maximum (+3.2 V vs. the normal hydrogen electrode) [47].

The coloration of MoO$_3$, whether induced by electrochemical proton injection (electrochromism) or catalytic hydrogen dissociation (gasochromism), is well studied. The protons produced, in conjunction with photogenerated electrons, facilitate the formation of molybdenum bronze H$_x$Mo$_x^{5+}$Mo$_{1-x}^{6+}$O$_3$, where hydrogen bonds with oxygen, enabling IVCT absorption between Mo$^{5+}$ and Mo$^{6+}$ sites [48]. The IVCT mechanism gives rise to a broad optical absorption band in the visible red to near-infrared region, typically between 600 and 900 nm [43], which contributes to the observed colour change and intensifies with increasing hydrogen concentration [49]. Simultaneously, the refractive index in this spectral range decreases from approximately 2.1 to 1.6 as the hydrogen content $x$ increases from 0.0 to 0.5 [49,50]. The overall coloration process in YHO/MoO$_3$ can be described by the following reactions:

$$YH_xO_y \xrightarrow{h\nu} YH_{x-z}O_y + \frac{z}{2} H_2 \uparrow \tag{1}$$

$$MoO_3 \xrightarrow{h\nu > E_g} MoO_3 + e^- + h^+ \qquad (2)$$

$$H_2 + 2h^+ \rightarrow 2H^+ \qquad (3)$$

$$MoO_3 + xH^+ + xe^- \rightarrow H_xMo_x^{5+}Mo_{1-x}^{6+}O_3 \qquad (4)$$

Additionally, in the gasochromic experiments, hydrogen molybdenum bronze is not the only reaction product; instead, a portion of the material transforms into $MoO_{3-x}$, resulting in the formation of oxygen vacancies and the release of water [51]. Moreover, when exposed to air, $MoO_{3-x}$ does not completely revert to its initial $MoO_3$ phase. Experimental studies have shown that, in contrast to $WO_3$, coloration in $MoO_3$ is less reversible [52] and strongly depends on the surrounding atmosphere [53,54]. The underlying mechanism responsible for this irreversibility remains a subject of debate. Several explanations have been proposed in the literature, including: (i) the irreversible reduction of $Mo^{6+}$ to $Mo^{4+}$ accompanied by the formation and release of water [55], and (ii) the trapping of localized water molecules within the material that cannot diffuse out [56]. However, in our case, no evidence of $Mo^{4+}$ was detected after illumination (see XPS results in Section 3.3), nor was there a significant increase in water-related absorption bands in the infrared spectrum.

For samples with varying $MoO_3$ film thickness (see Fig. 5(a)), a clear trend emerges: as the thickness of the $MoO_3$ film increases, the photo-darkening and overall contrast of the YHO/$MoO_3$ films become more pronounced, while they do not bleach back to the initial state. With increasing $MoO_3$ thickness, the volume of material available for hydrogen intercalation and reaction also increases, resulting in the formation of a greater quantity of molybdenum bronze or sub-stoichiometric $MoO_{3-x}$. This, in turn, enhances the contrast of photo-darkening in the YHO/$MoO_3$ structure.

An exception is observed for the YHO/$MoO_3$ sample with a 285 nm thick $MoO_3$ film, where the contrast after the first illumination is lower compared to samples with thinner $MoO_3$ films. This difference can be attributed to the fact that the samples are illuminated from the $MoO_3$ side. Light absorption measurements of single $MoO_3$ films (see Fig. S8(a) in the SI) show that absorption increases with film thickness, reaching approximately 50% at 3.3 eV for a 285 nm thick film. Consequently, the intensity of radiation reaching the YHO film is reduced, and, as reported in the literature, the rate of coloration in the YHO film depends on the intensity of UV radiation (i.e., the number of photons reaching the YHO film) [4,57]. The difference in transmittance after illumination between the YHO/MoO$_3$ sample with a 285 nm MoO$_3$ film and the other samples decreases in subsequent cycles. It is expected that this sample would exhibit the lowest transmittance among the samples after illumination in later cycles, as Fig. 4(a) clearly demonstrates that, after 20 hours of illumination, the film shows the highest contrast. As discussed previously, the coloration of MoO$_3$ is stable and does not readily revert to its original transparent state under ambient conditions. Therefore, the bleaching behaviour observed in the YHO/MoO$_3$ samples primarily originates from the YHO layer. The transmittance in the bleached state ($T_{bleached}$) after illumination cycles is influenced by the extent of H$_x$MoO$_3$ formation: less formation results in higher $T_{bleached}$.

For samples with varying deposition pressures of the MoO$_3$ film (see Fig. 5(b)), several factors influence the observed photo-darkening and bleaching behaviour. Firstly, the deposition pressure during magnetron sputtering generally influences the density of the deposited film – lower pressures typically lead to denser films, whereas higher pressures result in more porous structures. This trend is also observed in the present case, as shown in Fig. S9 (SI), where the refractive index increases with decreasing pressure, indicating densification, a behaviour also reported for WO$_3$ [58]. Porous MoO$_3$ films facilitate hydrogen intercalation and chromic reactions. However, since all samples in this set have MoO$_3$ films of approximately the same thickness (≈80 nm) on top of the YHO film, denser films inherently contain a larger amount of material that can react with hydrogen. This leads to films with greater light absorption after prolonged illumination (see Fig. 4(b)). Finally, denser MoO$_3$ films absorb more light at 3.3 eV (see Fig. S8(b) in the SI), which reduces the intensity of radiation reaching the YHO film and consequently slows the darkening process. The slight variation in the composition of the MoO$_3$ film with deposition pressure (Table S1 in the SI) may also influence the photochromic properties; however, its exact effect remains unclear, as none of the single MoO$_3$ films exhibited photochromic activity.

Previous studies have shown that heating YHO/WO$_3$ samples can promote bleaching [24]. However, in our experiments, heating the samples to 90°C did not result in any significant improvement in transparency. While lower temperatures may facilitate faster bleaching of the YHO film, as suggested in [1], they may not be sufficient to restore MoO$_3$ to its fully transparent state [59]. Higher temperatures might be required, but excessive heating could ultimately degrade the photochromic properties of the YHO film [4].

Additional YHO/MoO$_3$ samples (MoO$_3$: 400 nm, 1.35 Pa) were produced by pre-illuminating the YHO layer with UVA-violet light for 20 hours to induce darkening, allowing it to recover to its initial state, and then depositing the MoO$_3$ film. This was done to confirm that hydrogen release from YHO occurs primarily during the initial illumination. Fig. 6 shows the transmittance spectra of YHO and YHO/MoO$_3$, both with and without prior YHO illumination, in both their clear and darkened states. It is evident that when YHO is pre-illuminated, there is no longer a significant increase in photochromic contrast of YHO/MoO$_3$ after illumination, as the release of hydrogen into MoO$_3$ is greatly reduced.

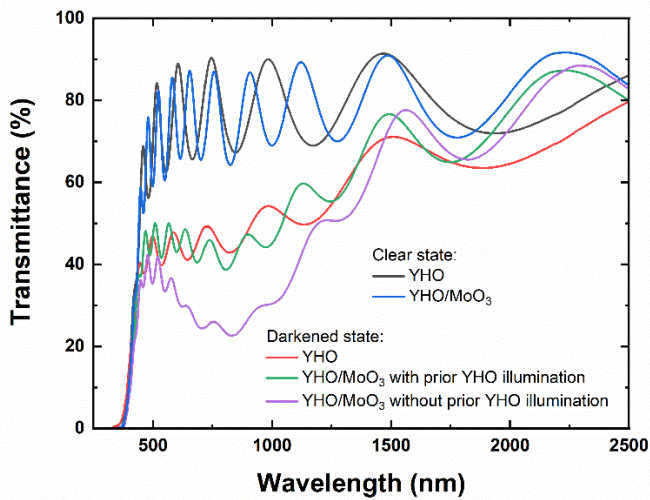

**Fig. 6.** Transmittance spectra in the 250–2500 nm range for single YHO and YHO/MoO$_3$, with and without prior illumination of YHO, in both clear and darkened states. A 400 nm MoO$_3$ layer was deposited at a pressure of 1.35 Pa with an O$_2$-to-Ar gas flow ratio of 1:3.

A significant portion of the photochromic contrast in YHO/MoO$_3$ still originates from the intrinsic photochromic properties of YHO. However, our interpretation does not consider possible modifications to these properties resulting from the presence of different types of MoO$_3$ cover layers, as this falls beyond the scope of the present study. The mechanisms underlying YHO photochromism in different environments are not yet fully understood. On the one hand, bleaching kinetics has been shown to be highly dependent on the surrounding environment [9,46]. On the other hand, single YHO films and those encapsulated with $Al_2O_3$ or $Si_3N_4$ diffusion barriers exhibit nearly identical photochromic properties, apart from the fact that encapsulated films demonstrate greater stability [60].

### 3.3. Composition and stability of YHO/MoO₃

The composition and chemical state of single $MoO_3$ and YHO/$MoO_3$ ($MoO_3$: 80 nm, 1.35 Pa) film surfaces were analysed using XPS. The survey spectra revealed well-defined signals from Mo and O elements. Based on the measured intensities, the O-to-Mo atomic % ratio was found to increase from 2.95 to 3.20 as the deposition pressure increased from 0.15 to 2.70 Pa (see Table S1 in the SI). This confirms that the films are close to stoichiometric composition. In the high-resolution spectra of the single $MoO_3$ films (Fig. 7(a)), the symmetric peaks corresponding to Mo $3d_{5/2}$ and Mo $3d_{3/2}$ were observed at 233.2 eV and 236.4 eV, respectively. These binding energies correspond to Mo in the +6 oxidation state [61].

In $MoO_3$ films, molybdenum can be partially reduced from its hexavalent state ($Mo^{6+}$) to pentavalent ($Mo^{5+}$) during the formation of either $H_xMoO_3$ or sub-stoichiometric $MoO_{3-x}$. The resulting increase in light absorption is explained by the $Mo^{5+}/Mo^{6+}$ IVCT model. XPS measurements allow quantification of these changes, and the results for both non-illuminated and two-hour illuminated single $MoO_3$ and bilayer YHO/$MoO_3$ films are presented in Fig. 7.

Fitting the high-resolution Mo 3d XPS spectra confirms that no reduction of Mo occurs in single $MoO_3$ films after two hours of UVA-violet illumination (Fig. 7(b)). This finding is consistent with the photochromic measurements, which showed that $MoO_3$ alone does not exhibit a photochromic effect. However, significant changes were observed in the $MoO_3$ film deposited on YHO. Before illumination, the $MoO_3$ film on YHO consisted of 89% $Mo^{6+}$ and 11% $Mo^{5+}$ (Fig. 7(c)). After two hours of illumination, the $Mo^{6+}$ fraction decreased to 47%, while the $Mo^{5+}$ fraction increased to 53% (Fig. 7(d)). The O-to-Mo ratio does not decrease after illumination, which rules out the formation of sub-stoichiometric $MoO_{3-x}$ accompanied by oxygen/water release and instead supports the formation of $H_xMoO_3$. The high-resolution O 1s XPS spectra (Fig. 7) reveal a primary peak at 531.1 eV and a less intense peak at 532.0 eV, attributed to OH bonds. The contribution of

OH bonds in the O 1s spectra is notably higher in the bilayer YHO/MoO$_3$ film compared to the single MoO$_3$ film, indicating the formation of new OH bonds upon hydrogen insertion into MoO$_3$.

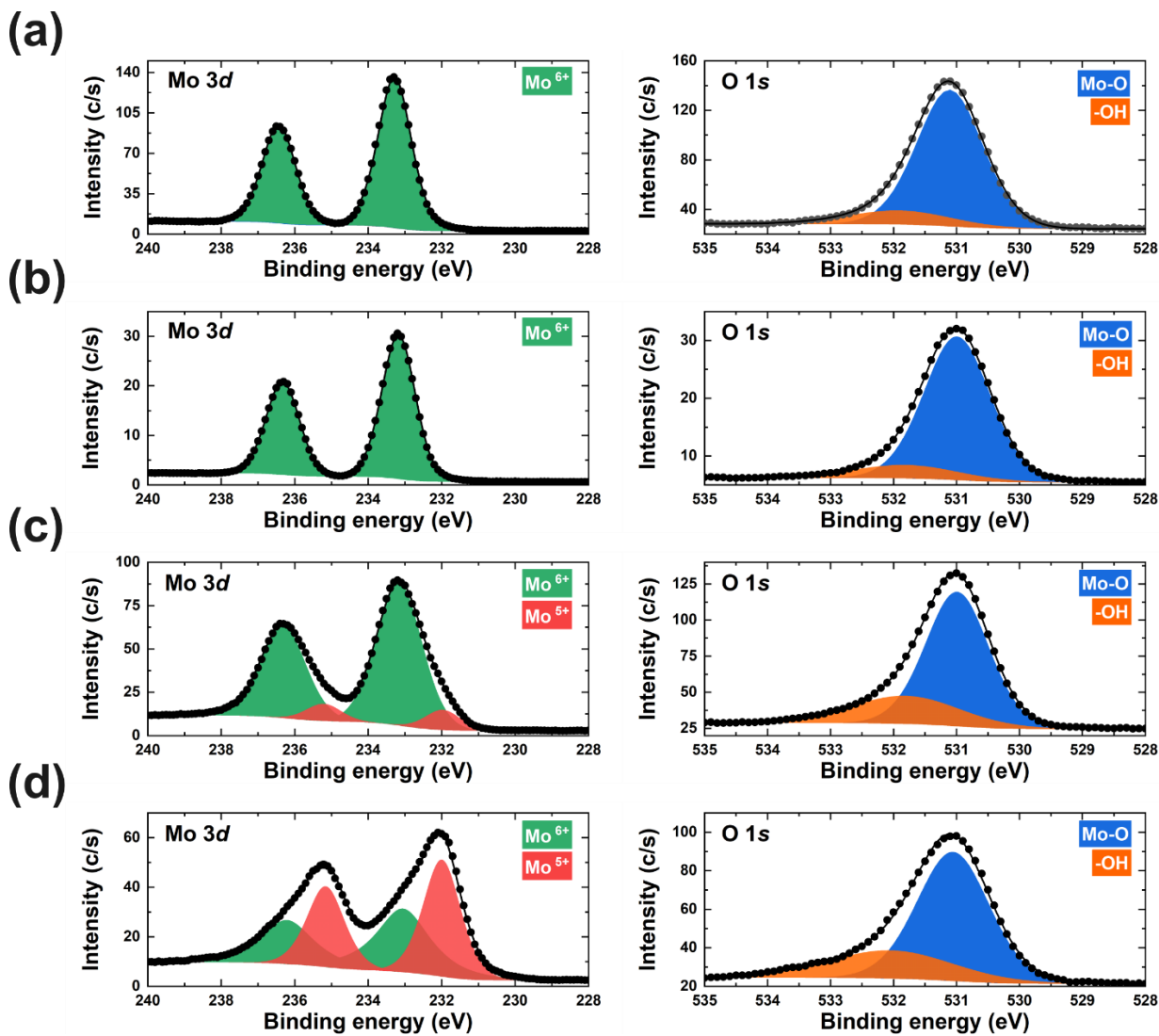

**Fig. 7.** High-resolution XPS spectra of Mo 3d and O 1s for (a) the non-illuminated and (b) the illuminated single MoO$_3$ film, and (c) the non-illuminated and (d) the illuminated bilayer YHO/MoO$_3$ film. The illumination was performed for two hours using UVA-violet light. In both cases, an 80 nm MoO$_3$ film was deposited at a pressure of 1.35 Pa with an O$_2$-to-Ar gas flow ratio of 1:3.

Even though exposure to light strongly enhances chromic reactions, the XPS results indicate that Mo reduction in the YHO/MoO$_3$ films progresses gradually, even in the absence of intentional illumination (Fig. 7(c)). Visually, some YHO/MoO$_3$ samples were observed to slowly develop a dark bluish tint over time, with the effect being more pronounced in samples with thicker MoO$_3$ films. Fig. 8(a) presents the transmittance spectra 44 days after deposition of a YHO/MoO$_3$ sample (MoO$_3$, 400 nm, 1.33 Pa) with and without prior YHO illumination, stored either in darkness or under ambient light conditions (exposed to natural daylight from a window and artificial room lighting). The key observation is that films stored in darkness retain higher transmittance values compared to those exposed to ambient light. Additionally, prior illumination of the YHO film influences the rate of transmittance change over time – samples that were pre-illuminated exhibit a slower colour change.

This phenomenon is attributed to the gradual release of hydrogen from the YHO film [19], which subsequently reacts with MoO$_3$ to form H$_x$MoO$_3$. To confirm this, we conducted composition measurements using NRA on samples without prior YHO illumination that were stored in darkness under an inert atmosphere. The NRA results reveal that the MoO$_3$ film deposited on YHO contains more than 10 at. % hydrogen, whereas MoO$_3$ deposited directly on glass contains less than 1 at. % (see Fig. 8(b)). The Y-to-O ratio in the single YHO film is 1.14. Assuming the YHO composition follows the formula YH$_{2-x}$O$_x$, the resulting composition would be approximately YH$_{1.12}$O$_{0.88}$.

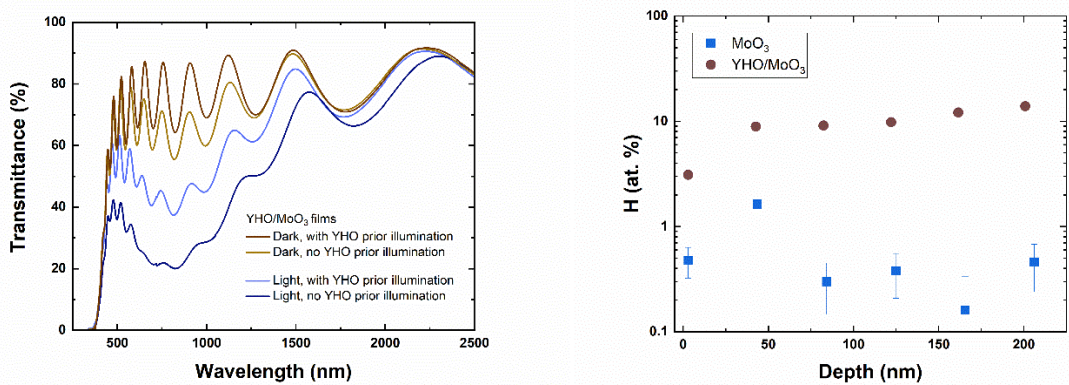

**Fig. 8.** (a) Transmittance spectra of YHO/MoO$_3$ (MoO$_3$, 400 nm, 1.33 Pa) after 44 days, comparing samples with and without prior YHO illumination, stored in darkness or under ambient light. (b) NRA depth profiles of hydrogen concentrations in the double-layered YHO/MoO$_3$ and single MoO$_3$ films.

## 4. Conclusions

In this study, YHO/MoO$_3$ composite thin films were synthesised using reactive magnetron sputtering, and their photochromic properties were systematically investigated in relation to MoO$_3$ thickness and density. The results confirm a synergistic effect between YHO and MoO$_3$, where the presence of MoO$_3$ enhances the photo-darkening response compared to that of a single YHO film. This enhancement is attributed to hydrogen intercalation from YHO into MoO$_3$, leading to the formation of H$_x$MoO$_3$. XPS analysis revealed a progressive reduction of Mo$^{6+}$ to Mo$^{5+}$ during UV illumination, further supporting the role of hydrogen migration in the photochromic mechanism.

The thickness and density of MoO$_3$ films played a crucial role in photochromic properties of YHO/MoO$_3$ structures. Films with thicker or denser MoO$_3$ films exhibited larger photochromic contrast after prolonged illumination due to increased hydrogen accommodation. The photo-darkened YHO/MoO$_3$ films did not fully recover to their original transparency, highlighting the persistent retention of hydrogen in the MoO$_3$ film. This behaviour differs from WO$_3$-based systems, which typically exhibit reversible colour changes. In contrast, the cumulative and less reversible darkening of YHO/MoO$_3$ films may be advantageous for passive optical applications such as disposable photo-exposure tags or indicators, where persistent visual change enables exposure tracking without the need for active power or switching mechanisms. Additionally, gradual darkening over time, even in the absence of intentional illumination, further confirms the diffusion-driven nature of the process. NRA provided direct evidence of significant hydrogen retention in MoO$_3$ when deposited on YHO, whereas single MoO$_3$ films contained negligible hydrogen. This result reinforces the hypothesis that hydrogen released from YHO is the primary factor driving the colouration of MoO$_3$.

Overall, these findings contribute to a deeper understanding of hydrogen dynamics in YHO/MoO$_3$ photochromic coatings and provide valuable insights for optimising deposition conditions to enhance optical performance. Future studies could focus on tailoring the YHO film – particularly its hydrogen content – to offer further control over hydrogen incorporation and release behaviour.

**Acknowledgements**

Financial support was provided by Latvian Council of Science Project No. lzp-2022/1-0454. The Institute of Solid State Physics, University of Latvia, as a Center of Excellence, has received funding from the European Union's Horizon 2020 Framework Programme H2020-WIDESPREAD-01-2016-2017-TeamingPhase2 under grant agreement No. 739508, project CAMART$^2$. The research team acknowledges SWEB project 101087367 funded by the HORIZON-WIDERA-2022-TALENTS-01. Accelerator operation at the Ion Technology Center national Swedish research infrastructure was supported by VR-RFI under contract #2019-00191.

**Data availability**

Data will be made available on request.

**References**

[1] T. Mongstad, C. Platzer-Björkman, J.P. Maehlen, L.P.A. Mooij, Y. Pivak, B. Dam, E.S. Marstein, B.C. Hauback, S.Z. Karazhanov, A new thin film photochromic material: Oxygen-containing yttrium hydride, Sol. Energy Mater. Sol. Cells 95 (2011) 3596-3599. https://doi.org/10.1016/j.solmat.2011.08.018.

[2] D. Moldarev, M.V. Moro, C.C. You, E.M. Baba, S.Z. Karazhanov, M. Wolff, D. Primetzhofer, Yttrium oxyhydrides for photochromic applications: Correlating composition and optical response, Phys. Rev. Mater. 2(11) (2018) 115203. https://doi.org/10.1103/PhysRevMaterials.2.115203.

[3] D. Moldarev, M. Wolff, E.M. Baba, M.V. Moro, C.C. You, D. Primetzhofer, S.Z. Karazhanov, Photochromic properties of yttrium oxyhydride thin films: Surface versus bulk effect, Materialia 11 (2020) 100706. https://doi.org/10.1016/j.mtla.2020.100706.

[4] C.C. You and S.Z. Karazhanov, Effect of temperature and illumination conditions on the photochromic performance of yttrium oxyhydride thin films, J. Appl. Phys. 128 (2020) 013106.https://doi.org/10.1063/5.0010132.

[5] S. Cornelius, G. Colombi, F. Nafezarefi, H. Schreuders, R. Heller, F. Munnik, B. Dam, Oxyhydride nature of rare-earth-based photochromic thin films, The journal of physical chemistry letters 10(6) (2019) 1342-1348. https://doi.org/10.1021/acs.jpclett.9b00088.

[6] H. Kageyama, K. Hayashi, K. Maeda, J.P. Attfield, Z. Hiroi, J.M. Rondinelli, K.R. Poeppelmeier, Expanding frontiers in materials chemistry and physics with multiple anions, Nature communications 9(1) (2018) 1-15. https://doi.org/10.1038/s41467-018-02838-4.

[7] C.C. You, D. Moldarev, T. Mongstad, D. Primetzhofer, M. Wolff, E.S. Marstein, S.Z. Karazhanov, Enhanced photochromic response in oxygen-containing yttrium hydride thin films transformed by an oxidation process, Solar Energy Materials and Solar Cells 166 (2017) 185-189. https://doi.org/10.1016/j.solmat.2017.03.023.

[8] B. Dam, F. Nafezarefi, D. Chaykina, G. Colombi, Z. Wu, S.W. Eijt, S. Banerjee, G. de Wijs, A. Kentgens, Perspective on the photochromic and photoconductive properties of Rare-Earth Oxyhydride thin films, Sol. Energy Mater. Sol. Cells 273 (2024) 112921. https://doi.org/10.1016/j.solmat.2024.112921.

[9] E.M. Baba, J. Montero, E. Strugovshchikov, E.Ö. Zayim, S. Karazhanov, Light-induced breathing in photochromic yttrium oxyhydrides, Physical Review Materials 4(2) (2020) 025201. https://doi.org/10.1103/PhysRevMaterials.4.025201.

[10] J. Montero, F.A. Martinsen, M. García-Tecedor, S.Z. Karazhanov, D. Maestre, B. Hauback, E.S. Marstein, Photochromic mechanism in oxygen-containing yttrium hydride thin films: An optical perspective, Physical review B 95(20) (2017) 201301. https://doi.org/10.1103/PhysRevB.95.201301.

[11] C.V. Chandran, H. Schreuders, B. Dam, J.W. Janssen, J. Bart, A.P. Kentgens, P.J.M. van Bentum, Solid-state NMR studies of the photochromic effects of thin films of oxygen-containing yttrium hydride, The Journal of Physical Chemistry C 118(40) (2014) 22935-22942. https://doi.org/10.1021/jp507248c.

[12] D. Moldarev, E. Pitthan, M. Wolff, D. Primetzhofer, Effects of H vacancies on photochromic properties of oxygen-containing yttrium hydride, Nucl. Instrum. Methods Phys. Res., B: Beam Interactions with Materials and Atoms 555 (2024) 165486. https://doi.org/10.1016/j.nimb.2024.165486.

[13] M. Hans, T.T. Tran, S.M. Aðalsteinsson, D. Moldarev, M.V. Moro, M. Wolff, D. Primetzhofer, Photochromic mechanism and dual-phase formation in oxygen-containing rare-earth hydride thin films, Advanced Optical Materials 8(19) (2020) 2000822. https://doi.org/10.1002/adom.202000822.

[14] J. Chai, S. Zewei, W. Han, M. Chen, O. Wanseok, Y. Tang, Z. Yong, Ultrafast processes in photochromic material YH x O y studied by excited-state density functional theory simulation, SCIENCE CHINA Materials 63(8) (2020) 1579-1587. https://doi.org/10.1007/s40843-020-1343-x.

[15] D. Chaykina, T. De Krom, G. Colombi, H. Schreuders, A. Suter, T. Prokscha, B. Dam, S. Eijt, Structural properties and anion dynamics of yttrium dihydride and photochromic oxyhydride thin films examined by in situ µ+ SR, Physical Review B 103(22) (2021) 224106. https://doi.org/10.1103/PhysRevB.103.224106.

[16] J. Montero, F.A. Martinsen, M. Lelis, S.Z. Karazhanov, B.C. Hauback, E.S. Marstein, Preparation of yttrium hydride-based photochromic films by reactive magnetron sputtering, Solar Energy Materials and Solar Cells, 177 (2018) 106-109. https://doi.org/10.1016/j.solmat.2017.02.001.

[17] M. Zubkins, I. Aulika, E. Strods, V. Vibornijs, L. Bikse, A. Sarakovskis, G. Chikvaidze, J. Gabrusenoks, H. Arslan, J. Purans, Optical properties of oxygen-containing yttrium hydride thin films during and after the deposition, Vacuum 203 (2022) 111218. https://doi.org/10.1016/j.vacuum.2022.111218.

[18] J. Montero, Photochromism in rare earth oxyhydrides for large-area transmittance control, Sol. Energy Mater. Sol. Cells 272 (2024) 112900. https://doi.org/10.1016/j.solmat.2024.112900.

[19] D. Moldarev, D. Primetzhofer, C.C. You, S.Z. Karazhanov, J. Montero, F. Martinsen, T. Mongstad, E.S. Marstein, M. Wolff, Composition of photochromic oxygen-containing yttrium hydride films, Sol. Energy Mater. Sol. Cells 177 (2018) 66-69. https://doi.org/10.1016/j.solmat.2017.05.052.

[20] M. Zubkins, J. Gabrusenoks, R. Aleksis, G. Chikvaidze, E. Strods, V. Vibornijs, A. Lends, K. Kundzins, J. Purans, Vibrational Properties of Photochromic Yttrium Oxyhydride and

Oxydeuteride Thin Films, Journal of Alloys and Compounds 1015 (2025) 178917. https://doi.org/10.1016/j.jallcom.2025.178917.

[21] S. Banerjee, D. Chaykina, R. Stigter, G. Colombi, S.W. Eijt, B. Dam, A. Gilles, A.P. Kentgens, Exploring Multi-Anion Chemistry in Yttrium Oxyhydrides: Solid-State NMR Studies and DFT Calculations, The Journal of Physical Chemistry C 127.29 (2023) 14303 14316. https://doi.org/10.1021/acs.jpcc.3c02680.

[22] G. Colombi, S. Cornelius, A. Longo, B. Dam, Structure model for anion-disordered photochromic gadolinium oxyhydride thin films, The Journal of Physical Chemistry C 124.25 (2020) 13541-13549. https://doi.org/10.1021/acs.jpcc.0c02410.

[23] M. La, N. Li, R. Sha, S. Bao, P. Jin, Excellent photochromic properties of an oxygen-containing yttrium hydride coated with tungsten oxide (YHx: O/WO$_3$), Scripta Materialia 142 (2018): 36-40. https://doi.org/10.1016/j.scriptamat.2017.08.020.

[24] Q. Zhang, L. Xie, Y. Zhu, Y. Tao, R. Li, J. Xu, S. Bao, P. Jin, Photo-thermochromic properties of oxygen-containing yttrium hydride and tungsten oxide composite films, Sol. Energy Mater. Sol. Cells 200 (2019) 109930. https://doi.org/10.1016/j.solmat.2019.109930.

[25] I.A. De Castro, R.S. Datta, J.Z. Ou, A. Castellanos-Gomez, S. Sriram, T. Daeneke, K. Kalantar-Zadeh, Molybdenum oxides–from fundamentals to functionality, Advanced Materials 29 (2017) 1701619. https://doi.org/10.1002/adma.201701619.

[26] H. Zheng, J.Z. Ou, M.S. Strano, R.B. Kaner, A. Mitchell, K. Kalantar-Zadeh, Nanostructured tungsten oxide–properties, synthesis, and applications, Advanced Functional Materials 21 (2011) 2175-2196. https://doi.org/10.1002/adfm.201002477.

[27] S. Santhosh, M. Mathankumar, S. Selva Chandrasekaran, A.K. Nanda Kumar, P. Murugan, B. Subramanian, Effect of ablation rate on the microstructure and electrochromic properties of pulsed-laser-deposited molybdenum oxide thin films, Langmuir 33 (2017) 19-33. https://doi.org/10.1021/acs.langmuir.6b02940.

[28] D.D. Yao, J.Z. Ou, K. Latham, S. Zhuiykov, A.P. O'Mullane, K. Kalantar-Zadeh, Electrodeposited α-and β-phase MoO$_3$ films and investigation of their gasochromic properties, Crystal growth & design 12 (2012) 1865-1870. https://doi.org/10.1021/cg201500b.

[29] M. Ranjbar, F. Delalat, H. Salamati, Molybdenum oxide nanosheets prepared by an anodizing-exfoliation process and observation of photochromic properties, Applied Surface Science 396 (2017) 1752-1759. https://doi.org/10.1016/j.apsusc.2016.11.225.

[30] S.A. Tomás, M.A. Arvizu, O. Zelaya-Angel, P. Rodríguez, Effect of ZnSe doping on the photochromic and thermochromic properties of MoO3 thin films, Thin Solid Films 518 (2009) 1332-1336. https://doi.org/10.1016/j.tsf.2009.05.054.

[31] C.S. Hsu, C.C. Chan, H.T. Huang, C.H. Peng, W.C. Hsu, Electrochromic properties of nanocrystalline MoO3 thin films, Thin Solid Films 516 (2008) 4839-4844. https://doi.org/10.1016/j.tsf.2007.09.019.

[32] T. He, Y. J. Yao, Photochromism of molybdenum oxide, Journal of Photochemistry and Photobiology C: Photochemistry Reviews 4.2 (2003) 125-143. https://doi.org/10.1016/S1389-5567(03)00025-X.

[33] P. Ström, D. Primetzhofer, Ion beam tools for nondestructive in-situ and in-operando composition analysis and modification of materials at the Tandem Laboratory in Uppsala, Journal of instrumentation 17 (2022) P04011. https://doi.org/10.1088/1748-0221/17/04/P04011.

[34] S.M. Aðalsteinsson, M.V. Moro, D. Moldarev, S. Droulias, M. Wolff, D. Primetzhofer, Correlating chemical composition and optical properties of photochromic rare-earth oxyhydrides using ion beam analysis, Nuclear Instruments and Methods in Physics Research Section B: Beam Interactions with Materials and Atoms 485 (2020) 36-40. https://doi.org/10.1016/j.nimb.2020.09.016.

[35] K. Gesheva, A. Szekeres, T. Ivanova, Optical properties of chemical vapor deposited thin films of molybdenum and tungsten based metal oxides, Solar energy Materials and Solar cells 76 (2003) 563-576. https://doi.org/10.1016/S0927-0248(02)00267-2.

[36] A. Šutka, M. Zubkins, A. Linarts, L. Lapčinskis, K. Mālnieks, O. Verners, A. Sarakovskis, R. Grzibovskis, J. Gabrusenoks, E. Strods, K. Smits, V. Vibornijs, L. Bikse, J. Purans, Tribovoltaic device based on the W/WO3 schottky junction operating through hot carrier extraction, The Journal of Physical Chemistry C 125 (2021) 14212-14220. https://doi.org/10.1021/acs.jpcc.1c04312.

[37] J.P. Maehlen, T.T. Mongstad, C.C. You, S. Karazhanov, Lattice contraction in photochromic yttrium hydride, Journal of alloys and compounds 580 (2013) S119-S121. https://doi.org/10.1016/j.jallcom.2013.03.151.

[38] T. Mongstad, C. Platzer-Björkman, S.Z. Karazhanov, A. Holt, J.P. Maehlen, B.C. Hauback, Transparent yttrium hydride thin films prepared by reactive sputtering, Journal of Alloys and compounds 509 (2011) S812-S816. https://doi.org/10.1016/j.jallcom.2010.12.032.

[39] C.C. You, T. Mongstad, E.S. Marstein, S.Z. Karazhanov, The dependence of structural, electrical and optical properties on the composition of photochromic yttrium oxyhydride thin films, Materialia 6 (2019) 100307. https://doi.org/10.1016/j.mtla.2019.100307.

[40] C.C. You, T. Mongstad, J.P. Maehlen, S. Karazhanov, Engineering of the band gap and optical properties of thin films of yttrium hydride, Applied physics letters 105(3) (2014) 031910. https://doi.org/10.1063/1.4891175.

[41] M.G. da Silva Júnior, L.C.C. Arzuza, H.B. Sales, R.M.D.C. Farias, G.D.A. Neves, H.D.L. Lira, R.R. Menezes, A Brief Review of $MoO_3$ and $MoO_3$-Based Materials and Recent Technological Applications in Gas Sensors, Lithium-Ion Batteries, Adsorption, and Photocatalysis, Materials 16 (2023) 7657. https://doi.org/10.3390/ma16247657.

[42] T.S. Sian and G.B. Reddy, Optical, structural and photoelectron spectroscopic studies on amorphous and crystalline molybdenum oxide thin films, Sol. Energy Mater. Sol. Cells 82 (2004) 375-386. https://doi.org/10.1016/j.solmat.2003.12.007.

[43] A. Gavrilyuk, U. Tritthart, W. Gey, The nature of the photochromism arising in the nanosized $MoO_3$ films, Sol. Energy Mater. Sol. Cells 95 (2011) 1846-1851. https://doi.org/10.1016/j.solmat.2011.02.006.

[44] M. Rouhani, Y.L. Foo, J. Hobley, J. Pan, G.S. Subramanian, X. Yu, A. Rusydi, S. Gorelik, Photochromism of amorphous molybdenum oxide films with different initial $Mo^{5+}$ relative concentrations, Applied surface science 273 (2013) 150-158. https://doi.org/10.1016/j.apsusc.2013.01.218.

[45] S.K. Deb, Opportunities and challenges in science and technology of $WO_3$ for electrochromic and related applications, Solar Energy Materials and Solar Cells 92 (2008) 245-258. https://doi.org/10.1016/j.solmat.2007.01.026.

[46] D. Moldarev, L. Stolz, M.V. Moro, S.M. Aðalsteinsson, I.A. Chioar, S.Z. Karazhanov, D. Primetzhofer, M. Wolff, Environmental dependence of the photochromic effect of oxygen-containing rare-earth metal hydrides, J. Appl. Phys. 129 (2021) 153101. https://doi.org/10.1063/5.0041487.

[47] Y. He, L. Zhang, X. Wang, Y. Wu, H. Lin, L. Zhao, W. Weng, H. Wan, M. Fan, Enhanced photodegradation activity of methyl orange over Z-scheme type MoO 3–gC 3 N 4 composite under visible light irradiation, RSC Advances 4 (2014) 13610-13619. https://doi.org/10.1039/C4RA00693C.

[48] Y.A. Yang, Y.W. Cao, B.H. Loo, J.N. Yao, Microstructures of electrochromic MoO3 thin films colored by injection of different cations, The Journal of Physical Chemistry B 102 (1998) 9392-9396. https://doi.org/10.1021/jp9825922.

[49] Z. Hussain, Optical constants and electrochromic characteristics of HxMoO3 and LixMoO3 bronzes, Journal of the Optical Society of America A 35.5 (2018) 817-829. https://doi.org/10.1364/JOSAA.35.000817.

[50] Z. Hussain, Dopant-dependent reflectivity and refractive index of microcrystalline molybdenum–bronze thin films, J. Appl. Phys. 91.9 (2002) 5745-5759. https://doi.org/10.1063/1.1461881.

[51] J.Z. Ou, J.L. Campbell, D. Yao, W. Wlodarski, K. Kalantar-Zadeh, In situ Raman spectroscopy of H2 gas interaction with layered MoO3, The Journal of Physical Chemistry C 115 (2011) 10757-10763. https://doi.org/10.1021/jp202123a.

[52] C. Park, S. Han, L.T. Duy, R. Yeasmin, G. Jung, D.W. Jeon, W. Kim, S.B. Cho, H. Seo, Gasochromic WO3/MoO3 sensors prepared by co-sputtering for hydrogen leak detection with long-lasting visible signal, Sensors and Actuators B: Chemical 426 (2025) 137030. https://doi.org/10.1016/j.snb.2024.137030.

[53] C. Bechinger, G. Oefinger, S. Herminghaus, P. Leiderer, On the fundamental role of oxygen for the photochromic effect of WO3, J. Appl. Phys. 74 (1993) 4527-4533. https://doi.org/10.1063/1.354370.

[54] M. Bourdin, G. Salek, A. Fargues, S. Messaddeq, Y. Messaddeq, T. Cardinal, M. Gaudon, Investigation on the coloring and bleaching processes of WO 3− x photochromic thin films, J. Mater. Chem. C 8 (2020) 9410-9421. https://doi.org/10.1039/D0TC02170A.

[55] A. Borgschulte, O. Sambalova, R. Delmelle, S. Jenatsch, R. Hany, F. Nüesch, Hydrogen reduction of molybdenum oxide at room temperature, Scientific reports 7.1 (2017) 40761. https://doi.org/10.1038/srep40761.

[56] S.S. Kalanur, I.H. Yoo, H. Seo, Pd on MoO3 nanoplates as small-polaron-resonant eye-readable gasochromic and electrical hydrogen sensor, Sensors and Actuators B: Chemical 247 (2017) 357-365. https://doi.org/10.1016/j.snb.2017.03.033.

[57] S. Kazi, D. Moldarev, M.V. Moro, D. Primetzhofer, M. Wolff, Correlating Photoconductivity and Optical Properties in Oxygen-Containing Yttrium Hydride Thin Films, Phys. Status Solidi RRL 17 (2023) 2200435. https://doi.org/10.1002/pssr.202200435.

[58] M. Zubkins, V. Vibornijs, E. Strods, I. Aulika, A. Zajakina, A. Sarakovskis, K. Kundzins, K. Korotkaja, Z. Rudevica, E. Letko, J. Purans, A stability study of transparent conducting WO3/Cu/WO3 coatings with antimicrobial properties, Surfaces and Interfaces 41 (2023) 103259. https://doi.org/10.1016/j.surfin.2023.103259.

[59] S. K. Deb, J. A. Chopoorian, Optical properties and color-center formation in thin films of molybdenum trioxide, J. Appl. Phys. 37.13 (1966) 4818-4825. https://doi.org/10.1063/1.1708145.

[60] M.V. Moro, S.M. Aðalsteinsson, T.T. Tran, D. Moldarev, A. Samanta, M. Wolff, D. Primetzhofer, Photochromic Response of Encapsulated Oxygen-Containing Yttrium Hydride Thin Films, Phys. Status Solidi Rapid Res. Lett. 15 (2001) 2000608. https://doi.org/10.1002/pssr.202000608.

[61] J. Swiatowska-Mrowiecka, S. De Diesbach, V. Maurice, S. Zanna, L. Klein, E. Briand, I. Vickridge, P. Marcus, Li-ion intercalation in thermal oxide thin films of MoO3 as studied by XPS, RBS, and NRA, J. Phys. Chem. C 112 (2008) 11050-11058. https://doi.org/10.1021/jp800147f.